\documentclass[aps,twocolumn,superscriptaddress,showpacs,amssymb,prb,floatfix,longbibliography]{revtex4-2}
\usepackage{graphicx }
\usepackage[]{hyperref}
\usepackage{amsmath}
\usepackage{amssymb}

\hypersetup{colorlinks=true,linkcolor=blue,citecolor=blue,urlcolor=blue,pdfpagemode=UseNone}

\begin{document}


\title{AC sensing using nitrogen vacancy centers in a diamond anvil cell up to 6 GPa}
\author{Z. Wang}
\author{C. McPherson}
\author{R. Kadado}
\author{N. Brandt}
\author{S. Edwards}
\affiliation{
Department of Physics and Astronomy, University of California, Davis CA 95616}
\author{W. H. Casey}
\affiliation{Department of Earth and Planetary Sciences, University of California, Davis, 95616}
\affiliation{Department of Chemistry, University of California, Davis, 95616}
\author{N. J. Curro}
\affiliation{
Department of Physics and Astronomy, University of California, Davis CA 95616
}%
 \email{curro@physics.ucdavis.edu.}

\date{\today}

\begin{abstract}
Nitrogen-vacancy color centers in diamond have attracted broad attention as quantum sensors for both static and dynamic magnetic, electrical, strain and thermal fields, and are particularly attractive for quantum sensing under pressure in diamond anvil cells. Optically-based nuclear magnetic resonance may be possible at pressures greater than a few GPa, and offers an attractive alternative to conventional Faraday-induction based detection.  Here we present AC sensing results and demonstrate synchronized readout up to 6 GPa, but find that the sensitivity is reduced due to inhomogeneities of the microwave field and pressure within the sample space.  These experiments enable the possibility for all-optical high resolution magnetic resonance of nanoliter sample volumes at high pressures.
\end{abstract}

\maketitle

\section{\label{sec:level1}Introduction}

Nuclear magnetic resonance is one of the most versatile tools for investigating condensed matter in extreme conditions, including temperatures ranging from sub-mK to 1000 K, static magnetic fields up to 45 T, pulsed magnetic fields \cite{HaasePulseFieldNMR}, uniaxial strain, and hydrostatic pressures up to  several GPa \cite{Meier_2018}.  Each of these environments presents special challenges for conducting magnetic resonance experiments, but pressure is exceptional because of the small volumes available. The maximum achievable pressure in of a given pressure cell scales roughly as the inverse of the sample volume, and depends on the material parameters and design of the cell.  Intermediate-scale pressures up to 2-3 GPa can be reached within a piston-cylinder type cell, with volumes on the order of several hundred $\mu$L.  For this type of pressure cell, it is straightforward to introduce a small coil and perform conventional NMR. This approach has become standard in the study of condensed matter systems \cite{Lin2015} as well as solution chemistry at high pressure \cite{Ochoa2015}.

\begin{figure}
\begin{center}
\includegraphics[width=\linewidth]{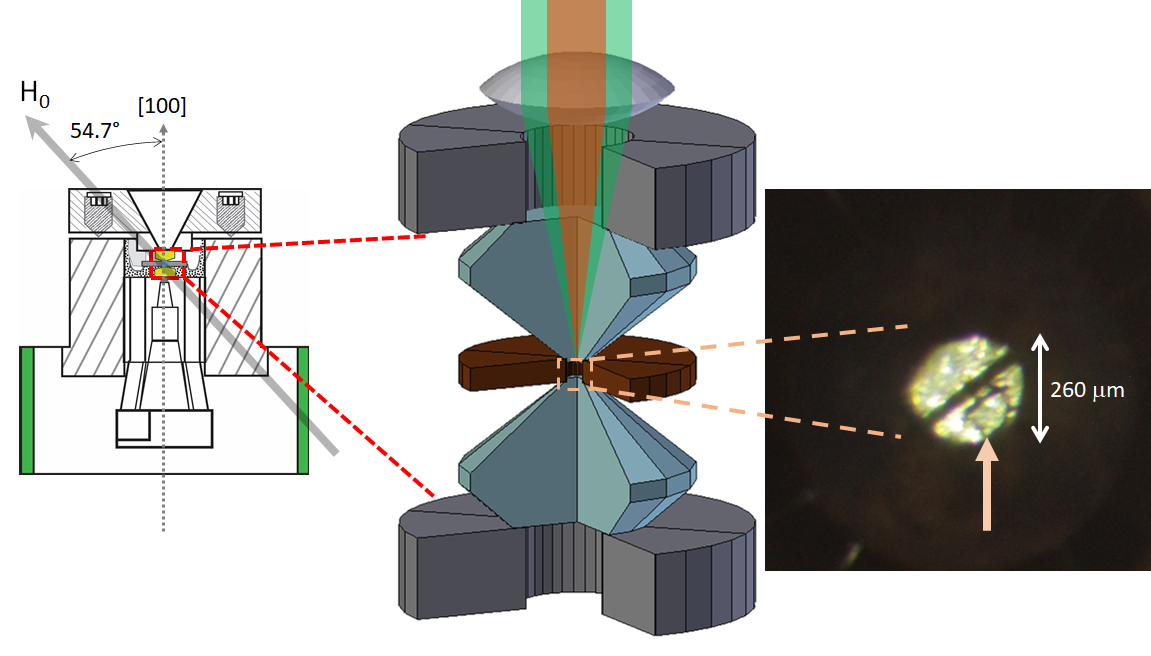}
\caption{Diamond anvil cell for optically-detected magnetic resonance experiments. An objective focuses the excitation light and collects the fluorescence  from within the sample space in the gasket. A  {straight} gold antenna is located across the anvil culet ( {sample hole 260} $\mu$m diameter), and a small NV diamond crystal of dimensions $\sim 100\times 75$ $\mu$m is visible  {at the center} to the lower right of the antenna  {(see arrow)}, oriented  {normal} the [100] direction. A magnetic field of magnitude $H_0 \sim 10$ mT is oriented $54.7^{\circ}$ from the vertical along the [111] direction of the diamond using permanent neodymium magnets mounted externally.}
\label{fig:dac}
\end{center}
\end{figure}

Piston-cylinder type cells  are limited to approximately 2-3 GPa due to the Young's modulus of the Cu-Be alloy that forms the cylinder housing \cite{Fujiwara1980}.  Alternative designs can extend this pressure range up to 4.5 GPa \cite{eremets}, although with severe constraints on the sample volume.  To reach pressures beyond this range, it is necessary to use a diamond anvil cell(DAC), in which a pair of diamond anvils compress a sample volume contained within a metallic gasket, as illustrated in Fig. \ref{fig:dac} \cite{eremets1996high}.
The available sample volume is only on the order of 0.1-1 nL, presenting serious challenges for conventional NMR. Not only should a solenoid  be located within this space to form a high $Q$ resonant circuit and introduce the necessary radiofrequency fields for NMR, but the leads must also be insulated from the gasket.  Preventing leads from shorting or breaking under pressure is a major challenge \cite{HaaseJMR2015}.  Secondly, the nuclear magnetization needs to be large enough to induce a sufficient voltage relative to the Johnson noise of the coil.  Typically this requires on the order of $10^{17}$ nuclear spins \cite{hoult1976signal}.  For a typical solid, e.g. copper with density 10 g/cm$^{3}$, this limit corresponds to 1 nL. Since the sample space must also contain the coil and pressure medium, there is often insufficient room for the minimum detectable sample.  Although there have been  {recent advances} to overcome these limitations using specialized microcoils, gasket resonators and Lenz lenses  {enabling NMR measurements up to the  {hundred GPa} regime} \cite{Meier2018,Meier_2018,Meier2019},
alternative detection methods may be advantageous in order to investigate matter at higher pressures, for example superconductivity in  {hydride compounds}(hydrogen-rich materials) \cite{Drozdov2015,Drozdov2019,Somayazulu2019}.

Optically-detected magnetic resonance (ODMR) of NV centers offers an alternative  {and complementary} approach that is not limited by the small sample volume  {and can detect nuclear spins within volumes as small as $10^{-3}$ nL} with sub-mHz resolution.  Several excellent reviews are available outlining the basics of this technique \cite{Doherty20131,Matsuzaki2016,Bucher2019}.  The fluorescence intensity of the NV centers is a strong function of the NV spin ($S=1$) orientation, and the NV centers can be optically pumped to the ground state with high polarization.  Single NV detection is possible with polarization $>99\%$ using optical pumping resonant with the zero-phonon line \cite{Childress2013}. For NV ensembles with non-resonant pumping using  532 nm light, the optical contrast is a few percent.  Because detection is optical-based rather than using Faraday induction, it is no longer necessary to locate a high Q resonant circuit  within the sample space.  The spin orientation can be measured optically, in which excitation  and fluorescence  light passes through the transparent diamond anvils and can be focused down to a point on the sample of size $\sim 1\mu$m, enabling work with small samples. ODMR has been utilized to study static strain and magnetization as a function of pressure and temperature \cite{Lesik2019,Hsieh2019,Shang2019, Yip2019}. {Here we describe AC sensing experiments using NV centers in a diamond anvil cell to probe dynamic fields, such as those arising from precessing nuclear spins.
In principle there is no upper limit for such measurements, and we discuss some of the challenges that emerge under pressure as the properties of the NV centers evolve.  }


\section{Apparatus}

The principle challenges to performing ODMR in a DAC are associated with the collecting sufficient fluorescence from the sample as well as introducing microwaves with sufficient homogeneity and magnitude. Fortunately, ODMR is similar in practice to Raman spectroscopy, which is a well-established technique under pressure in DACs.

\begin{figure}
\begin{center}
\includegraphics[width=\linewidth]{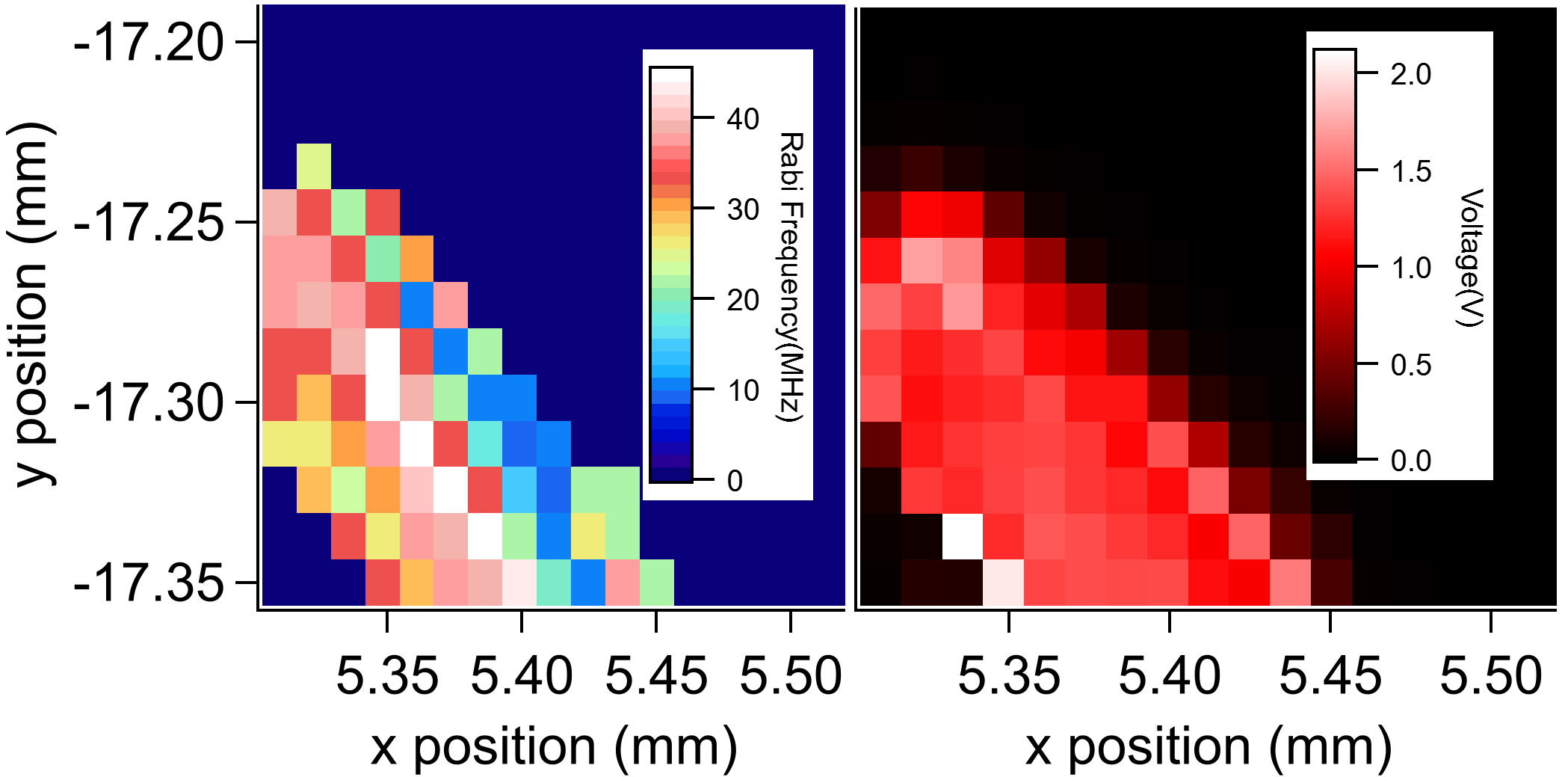}
\caption{(Left) Rabi frequency  {of the $|0\rangle \leftrightarrow |-1\rangle$ transition} and  (Right) fluorescence intensity measured as a function of position in an diamond crystal with NV centers located within a DAC.  {The scanning step size is {20 $\mu$m}}, or the scale of the individual squares on the map. }
\label{fig:map}
\end{center}
\end{figure}

\subsection{Design and Microwave Antenna}

Compared to ambient conditions, the signal-to-noise for ODMR in a DAC will be reduced because the structure of the cell prevents the objective from being too close to the sample.  For our design, illustrated in Fig. \ref{fig:dac}, the distance to the diamond anvil is  approximately 10 mm, plus 2.5 mm for the the anvil itself.  It is thus necessary to utilize a long-working distance objective (Nikon CFI T Plan Epi SLWD 50X Objective, N.A. 0.40 and working distance 22 mm). The pressure cell (CryoDAC SC, Almax-Easylab), is mounted on a translation stage and an external magnetic field, $\mathbf{H}_0$,  {of 29 mT} oriented at an angle of 54.7$^{\circ}$ from the vertical is introduced via a pair of fixed neodynium permanent magnets.  The gaskets are fabricated with 0.3 mm thick copper-beryllium, and the sample holes of diameter of 260 $\mu$m are drilled through the center of the pre-indented region (depth ranges from 100 to 150 $\mu$m) using an electrical discharge machine.  A small diamond crystal prepared with 0.3 ppm NVs (Element 6) was cut and polished to a thickness of 20 $\mu$m  (Applied Diamond/Diamond Delaware Knives), and then secured to the anvil with  {superglue such that} the $[100]$ axis is aligned parallel to the DAC axis.  The ODMR spectrum is measured as the cell is rotated  {along the DAC axis} in order to align the field along the $[111]$ direction parallel to the NV axis.  The diamond anvils are type IIac with low nitrogen impurities in order to minimize background fluorescence.  The spot size of the laser is  {less than 20 $\mu$m}, and can be scanned laterally across the sample space.  Representative scans of the Rabi frequency, $\Omega$, and fluorescence intensity are shown in Fig. \ref{fig:map}.

\begin{figure*}
\begin{center}
\includegraphics[width=\linewidth]{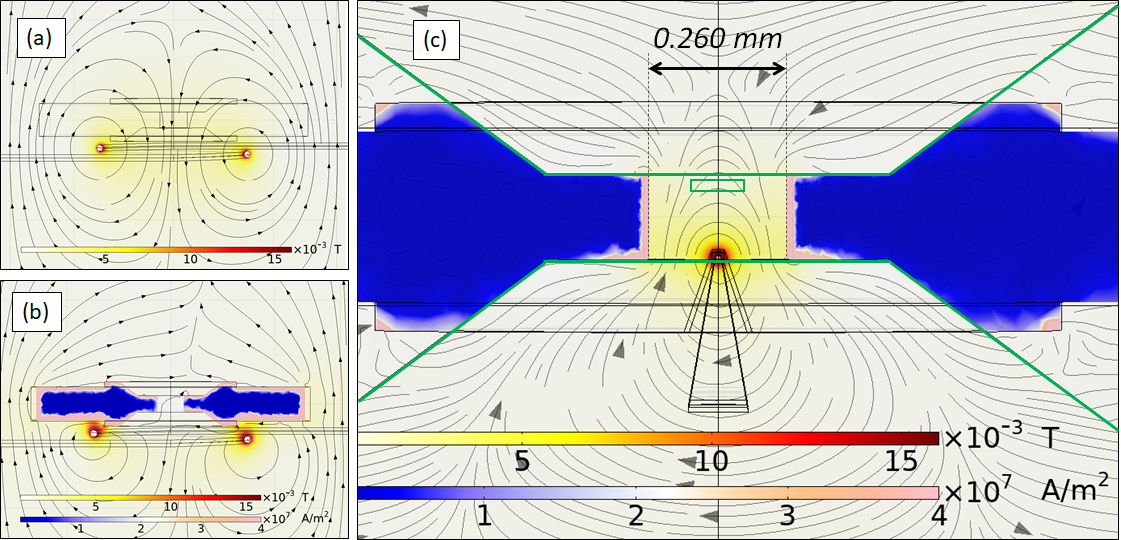}
\caption{ {Calculated magnetic field profile from a wire loop surrounding a diamond anvil close to (a) an insulating gasket and (b) a conducting gasket for 63.2 V(40 dBm) across the antenna at 3 GHz.  The conducting gasket screens the majority of the magnetic field within the sample space. (c) The calculated magnetic field profile and induced current density for a conducting gasket with a straight antenna, as illustrated in Fig. \ref{fig:antenna}(a). {The diamond anvil (outlined in green) has a relative permeability of 5 \cite{ibarra1997wide}.}  The green rectangle in the sample space represents the volume of the NV diamond chip used to compute the histograms in Fig. \ref{fig:Rabi}. } }
\label{fig:screening}
\end{center}
\end{figure*}

Introducing microwaves with sufficient magnitude in close proximity to the NV centers presents a major challenge for working in a DAC, because the conducting gasket can screen out the magnetic field, as illustrated in Fig. \ref{fig:screening}. High microwave amplitudes are necessary in order to achieve strong $\mathbf{H}_1$ fields and hence large Rabi frequencies.  The magnitude of $\mathbf{H}_1$ should be $\sim 0.18$ mT in order to have a 90$^{\circ}$ pulse time on the order of 50 ns ( {$\Omega \sim 5$} MHz). {Note that it is not necessary to have a high $Q$ resonant circuit, so the antenna does not necessarily need to have a low resistance.} Previous studies have utilized antennas located external to the sample chamber \cite{NVpressurePRL,Lesik2019,Hsieh2019,Yip2019}, within the sample chamber \cite{Shang2019}, as well as designer anvils with conducting paths deposited directly onto the diamond culet and protected with a capping layer of CVD grown diamond \cite{NVinDAC}.  Locating the antenna outside the sample space, as shown in Fig. \ref{fig:screening}(a,b), is easier, but is inefficient because a large fraction of the power goes towards inducing screening currents in the conducting gasket and can lead to undesired heating effects.  Designer anvils are costly and time consuming to prepare \cite{WeirDesignerDiamondLetter}.

\begin{figure}
\begin{center}
\includegraphics[width=\linewidth]{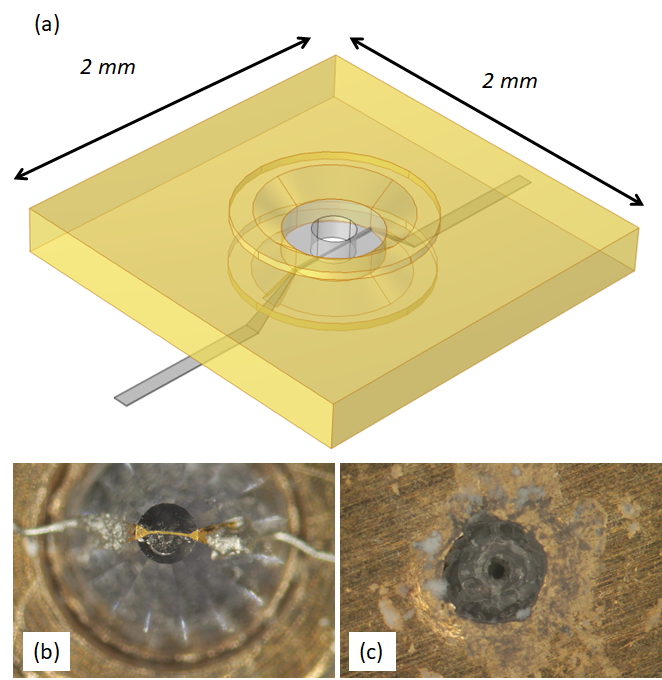}
\caption{ {(a) Geometry of the gasket and sample region.  The inner diameter is 260 $\mu$m, the culet diameter is 600 $\mu$m, and  the thickness of the sample region is 150 $\mu$m. The gasket is copper-beryllium with conductivity 1.16$\times 10^7$ S/m.  The gold antenna has width 11$\mu$m, thickness 8 $\mu$m and length 600 $\mu$m, and has conductivity 4.6$\times 10^7$ S/m.  There is also a thin insulating layer between the antenna and gasket with conductivity $10^{-18}$ S/m. (b) Fabricated microwave antenna secured onto culet of the anvil (diameter 600 $\mu$m). The antenna has width 10 $\mu$m and thickness  8 $\mu$m. (c) Pre-indented gasket with insulating layer of boronitride, aluminum oxide and epoxy.  The sample space in the center hole has diameter 260 $\mu$m.} }
\label{fig:antenna}
\end{center}
\end{figure}

\begin{figure*}
\begin{center}
\includegraphics[width=\linewidth]{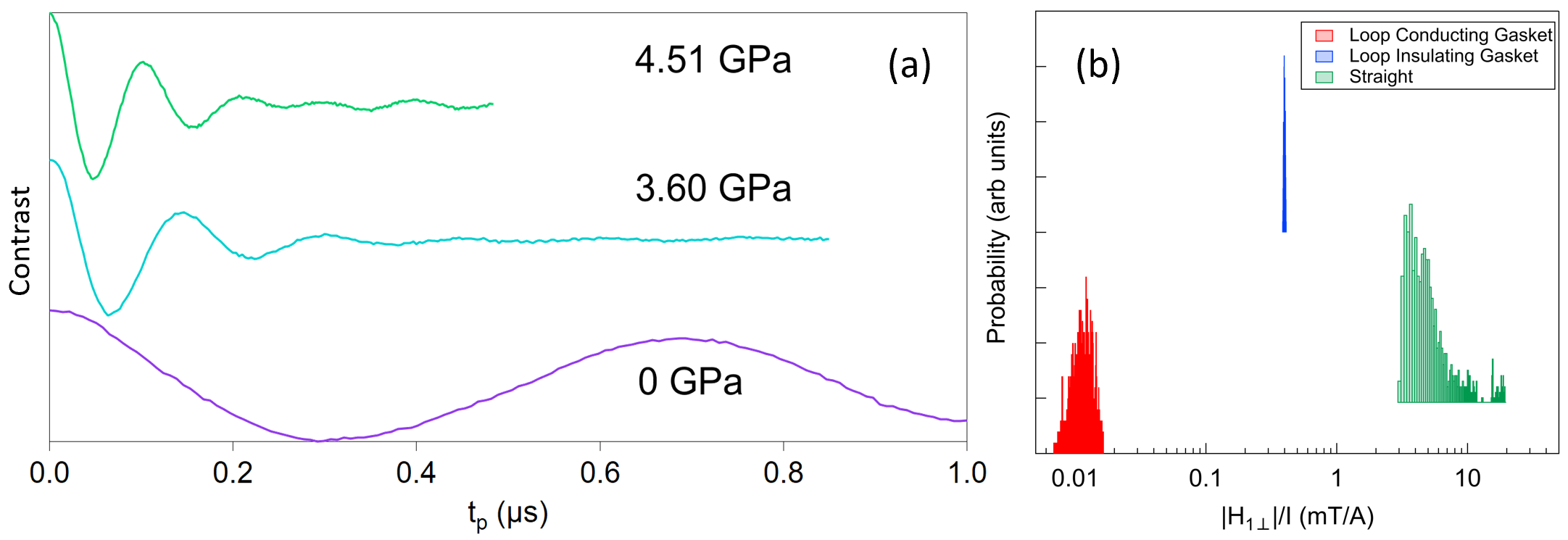}
\caption{(a) Rabi oscillations ( {$|0\rangle \leftrightarrow |-1\rangle$}) for various pressures in the DAC {with a microwave power of 44 dBm}.  (b)  Computed histograms of   $H_{1\perp}/I$ based upon the finite element simulations shown in Fig. \ref{fig:screening}(c),  for a loop antenna with insulating gasket, with a conducting gasket, and for the straight antenna design illustrated in Fig. \ref{fig:antenna}(a). Note that the field has been normalized by the current in the wire and have been computed at 2 GHz.  }
\label{fig:Rabi}
\end{center}
\end{figure*}

To overcome this challenge, we fabricated gold microwave strips, as shown in Fig. \ref{fig:antenna},  by electroplating 8 $\mu$m thick strips onto a substrate, which were then liberated chemically for insertion into the DAC.  A single antenna is then transferred onto the culet of the anvil with tweezers under a microscope, and secured with thin layer of adhesive.  Leads were attached to the ends of the antenna pads with silver epoxy, and then attached to larger wires (not shown) that lead out of the cell and to a high power (16 W)  microwave amplifier (Mini-Circuits, ZHL-16W-43-S+).  The antenna is insulated from the pre-indented region of the gasket with a mixture of epoxy and aluminum oxide and(or) boron nitride powders, that are cured under compression (Fig. \ref{fig:antenna}(c)).  {After the epoxy has cured, the center hole is drilled out, leaving a thin layer of insulating material between the conducting gasket and the antenna leads \cite{Meier2015}.} This approach enables us to reliably apply pressure to the gasket without comprising the antenna performance.  {The antenna leads are malleable and are compressed under pressure as the anvil presses into the gasket. However, at higher pressures (on the order of $\sim 10-30$ GPa) the leads will eventually sever as the gasket deforms and the anvil cups inward \cite{Meier2017a}. In such cases it may be better to utilize alternative designs, with thinner leads or even eliminating the leads altogether and using the Lenz lens approach of Ref. \cite{Meier2017a} For pressures up to $\sim 7$ GPa} we can achieve Rabi frequencies of several tens of MHz with our antenna, which correspond to $90^{\circ}$ pulses of a few tens of nanoseconds, as illustrated in Fig. \ref{fig:Rabi}.  { Eventually for sufficiently high pressure the antenna leads are thinned out and clipped by the anvils, limiting the power delivered and dramatically decreasing the Rabi frequency.} Microwaves pulse sequences are generated via an arbitrary waveform generator (Tabor AWG Model SE5082).

\subsection{Microwave field simulation}

In order to better understand the $\mathbf{H}_1$ field distribution and radiofrequency screening effects of the gasket, we have modeled the antenna and gasket system and carried out finite element analysis calculations of the electromagnetic fields.  Fig. \ref{fig:antenna}(a) illustrates the gasket and antenna strip, including the ridge surrounding the indented region around the anvils (not shown).  A thin insulating gap separates the antenna from the gasket.  Using the COMSOL package, we find that the field radiates radially from the anvil, as illustrated in Fig. \ref{fig:screening}(c), but drops off quickly with distance.  The ratio of power dissipated to resistance in the antenna to the radiated power is 3.2\%, and that the power dissipated by induced currents in the gasket is 0.5\%. These numbers indicate that the thin gold wire radiates power into the sample space efficiently.  On the other hand, for a loop antenna located outside the sample space (Fig. \ref{fig:screening}(b)), but in close proximity to the gasket and anvil, the magnetic field in the sample space is well-screened by the conducting gasket.   {In practice, the power delivered to the antenna is dominated by the reflectance coefficient of the combination of leads and antenna, and depends critically on the geometry of the leads external to the pressure cell.  The best practice is to keep the leads parallel to one another as much as possible.}

As seen in Fig. \ref{fig:screening}, the $\mathbf{H}_1$ field profile is inhomogeneous.  It is largest in the region closest to the antenna but drops off quickly along the vertical axis.  Fig. \ref{fig:Rabi}(b) displays a histogram of the field to current ratio of  magnitude of $\mathbf{H}_{1\perp}$, the component of $\mathbf{H}_1$ perpendicular to $\mathbf{H}_0$ along the [111] direction (see Fig. \ref{fig:dac}), within a volume of  $100\times 75 \times20$ $\mu$m$^3$ representing the NV diamond  {(illustrated by the green box in Fig. \ref{fig:screening}(c))}. The size and shape of this histogram depends critically on the location of the NV diamond within the sample space.  This distribution of fields means that each of the $\sim 8\times 10^9$ NV centers in the NV diamond experiences a slightly different Rabi frequency.  As a result, the Rabi oscillations fall off quickly, as observed in Fig. \ref{fig:Rabi}(a).
 {Experimentally, the oscillations die out faster under pressure, as observed in Fig. \ref{fig:Rabi}(a). This behavior reflects inhomogeneous broadening, as discussed below.  We also considered a loop, rather than a straight antenna, as shown  in Fig. \ref{fig:Rabi}(b).  If the gasket remains insulating, the distribution remains narrow. For the conducting gasket, the distribution broadens and is shifted to lower fields because the gasket efficiently screens the field. The highest field to current ratios are achieved with the straight antenna because the distance to the sample is smallest in this case, however the field is inhomogeneous.}
Experimentally, we find that the  {Rabi oscillations are quickly damped (implying a wider distribution of $H_{1\perp}$ fields)} when the sample is located directly on top of the antenna, but is more homogeneous when the sample is located on the opposite anvil.  The latter is advantageous for AC detection because a distribution of $H_1$ fields will give rise to a distribution of rotation angles for the 180$^{\circ}$ dynamical decoupling pulses.


\section{AC-ODMR}

NV centers can sense magnetic fields with a sensitivity of the order of 30 pT/$\sqrt{\rm{Hz}}$, which is sufficient to detect the dipolar magnetic field generated by a nuclear spin that is within a few $\mu$m of the NV center.  Although  {couplings to such spins are} not large enough to resolve in the NV spectrum, they can be detected via AC methods \cite{degen2017quantum,Glenn2018,Schmitt2017}. AC-ODMR can discern the amplitude $B_{ac}$, frequency, $\omega_{ac}$, and phase, $\phi_{ac}$, of a time-varying field oriented along the direction, $\hat{z}$, of the applied field, $\mathbf{H}_0$:
\begin{equation}
    \mathbf{B}_{ac}(t) = B_{ac}\hat{z}\sin(\omega_{ac}t + \phi_{ac}).\label{eq:1}
\end{equation}
There are several technical components necessary to successfully detect such a signal.  First, microwave pulses to efficiently manipulate the NV spin orientation are necessary, with Rabi frequencies $\Omega = \gamma H_{1\perp}$ of at least 5 MHz. If the microwave amplitude is too small, the pulses used to create unitary rotations of the NV spin will be too long and will create an upper bound on the range of detectable frequencies. Secondly, in order to `imprint' the nuclear spin coupling onto the phase of the NV centers, the NV ensemble must have a sufficiently low density such that dipolar coupling between NV centers does not severely limit the decoherence time ($T_2\sim 200$ $\mu$s in our case).  This condition provides an effective lower bound on the range of frequencies that can be detected.

\subsection{Quadrature Detection and Phase Cycling}

\begin{figure}
\begin{center}
\includegraphics[width=\linewidth]{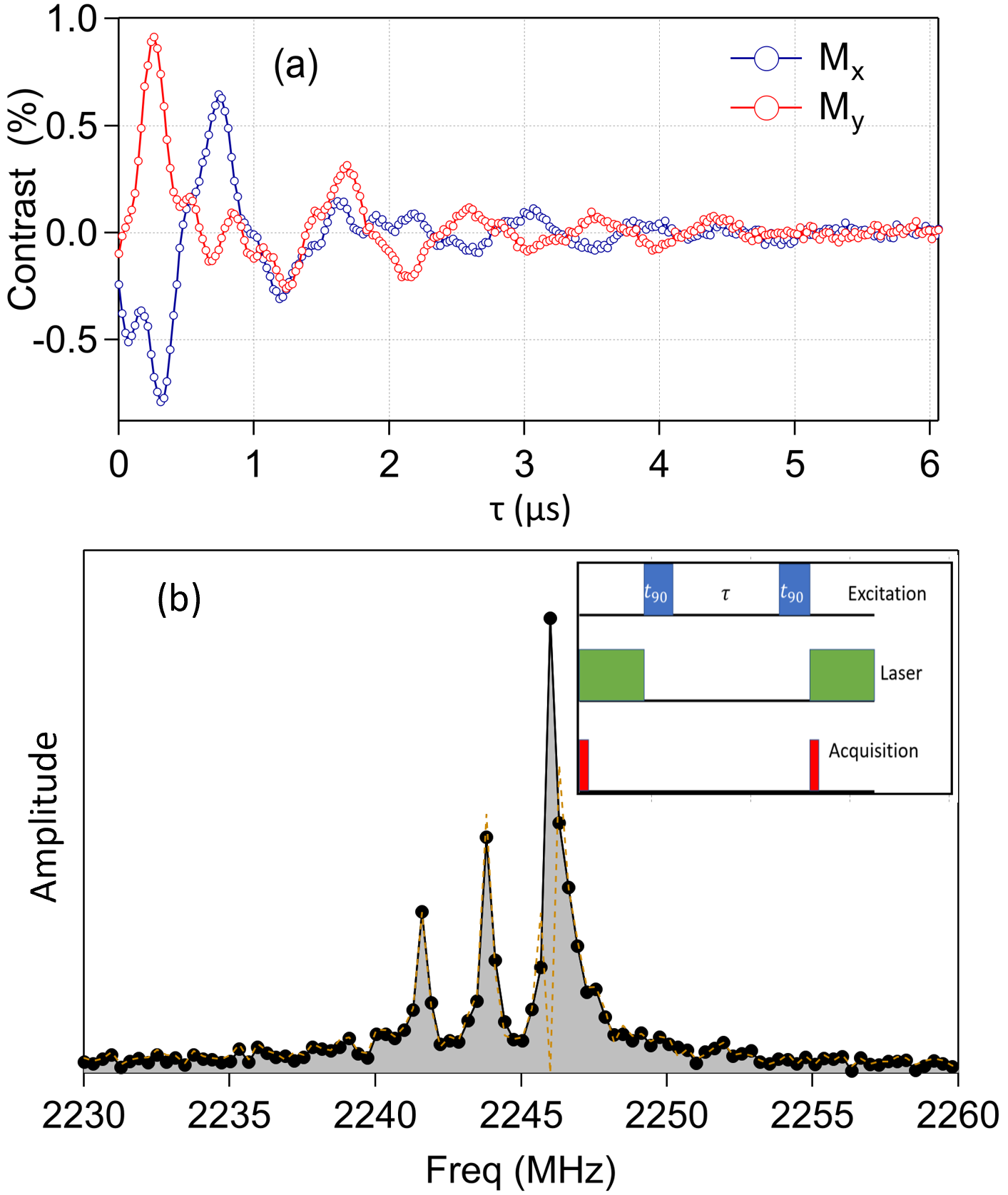}
\caption{(a) Change in fluorescence contrast as a function of pulse spacing, $\tau$, for a Ramsey sequence at the  {$|0\rangle \leftrightarrow |-1\rangle$} transition of an NV in a field of 22 mT  {at ambient pressure}.  $M_x$ and $M_y$ are measured using different phases (90$^{\circ}$)  for the second microwave pulse. Panel (b) shows the amplitude of the complex Fourier transform of the data in (a), revealing three separate peaks corresponding to the three $I_z$ states of the $^{14}$N nuclei coupled to the NV electronic spin.  {The dashed brown line shows the spectrum measured without phase cycling.}  }
\label{fig:quadrature}
\end{center}
\end{figure}

Conventional NMR relies on detecting the voltage induced by precessing spins in the plane perpendicular to the applied field, $\mathbf{H}_0$. In contrast, ODMR detects magnetization parallel to $\mathbf{H}_0$ ($\parallel \hat{z}$).  The fluorescence intensity, $I$, is a maximum, $I_0$, when the NV spins are polarized in the $S_z = 0$ state, and a minimum, $I_1$, for spins in the $S_{z} = \pm1$ states. The fluorescence contrast, $\Delta I = (I_0-I_1)/I_0$, is typically on the order of a few percent at room temperature, depending on the laser power and NV density.  Importantly, this contrast is proportional to $\langle S_z\rangle$.  Exposing the NVs to 532 nm excitation light for a few microseconds is sufficient to polarize the $S_z=0$ state, which can then be manipulated using oscillating microwave fields at the frequency of the $|0\rangle \leftrightarrow|-1\rangle$ or $|0\rangle \leftrightarrow|+1\rangle$ transitions. A $90^{\circ}$  $\mathbf{H}_1$ pulse will create a superposition state that will evolve with time and $\langle \mathbf{S}(t)\rangle$ will precess in the perpendicular plane. To detect the signal, one applies a second $90^{\circ}$ pulse with a particular phase (e.g. along either the $x$ or $y$ directions in the rotating frame) to rotate the magnetization back to the $z$ direction.  The so-called Ramsey sequence (Fig. \ref{fig:quadrature}(b) inset) will give rise to oscillations of $I/I_0$ as a function of the pulse spacing, $\tau$, between the two microwave pulses .

By controlling the phase of the second pulse relative to the first, one can project either $\langle S_x\rangle$ or $\langle S_y\rangle $ onto the $z$ axis.  Collecting data for both phases enables quadrature detection and hence the full complex Fourier transform can be obtained.   It is also useful to implement phase cycling to improve the signal-to-noise ratio.  Spurious noise, misalignments of the microwave pulses, and astigmatism in the phase response of the microwave sources can all contribute to noise in the measured fluorescence contrast,  {and similar approaches are regularly used in conventional magnetic resonance \cite{Bain1984}}. By cycling the phases of the first and second $\mathbf{H}_1$ pulses so that the NV spin is rotated alternately to the $\pm x$ and $\pm y$ directions and back to the $\pm z$ direction, one can either add or subtract the measured contrast to remove background noise.    Fig. \ref{fig:quadrature} demonstrates this approach.  Note that the spectrum of the NV reveals three separate peaks, which arise due to the hyperfine coupling between the NV spin, $S$, and the $^{14}$N nuclear spin, $I =1$ (natural abundance 99.6\%).  The hyperfine coupling is given by $\mathcal{H}_{hf} = \hat{\mathbf{S}}\cdot \mathbb{A}\cdot\hat{\mathbf{I}}$, where the coupling tensor $\mathbb{A}$ is diagonal along the NV-axis with components $A_{||} =2.1 $ and $A_{\perp} = 2.3$ MHz \cite{N14hyperfine}. The three peaks correspond to NV spins in the ensemble for which $I_z =$ -1, 0 or +1 and hence a hyperfine field of 0, $\pm 2.1$MHz shifts the resonant frequency.

\begin{figure}
\begin{center}
\includegraphics[width=\linewidth]{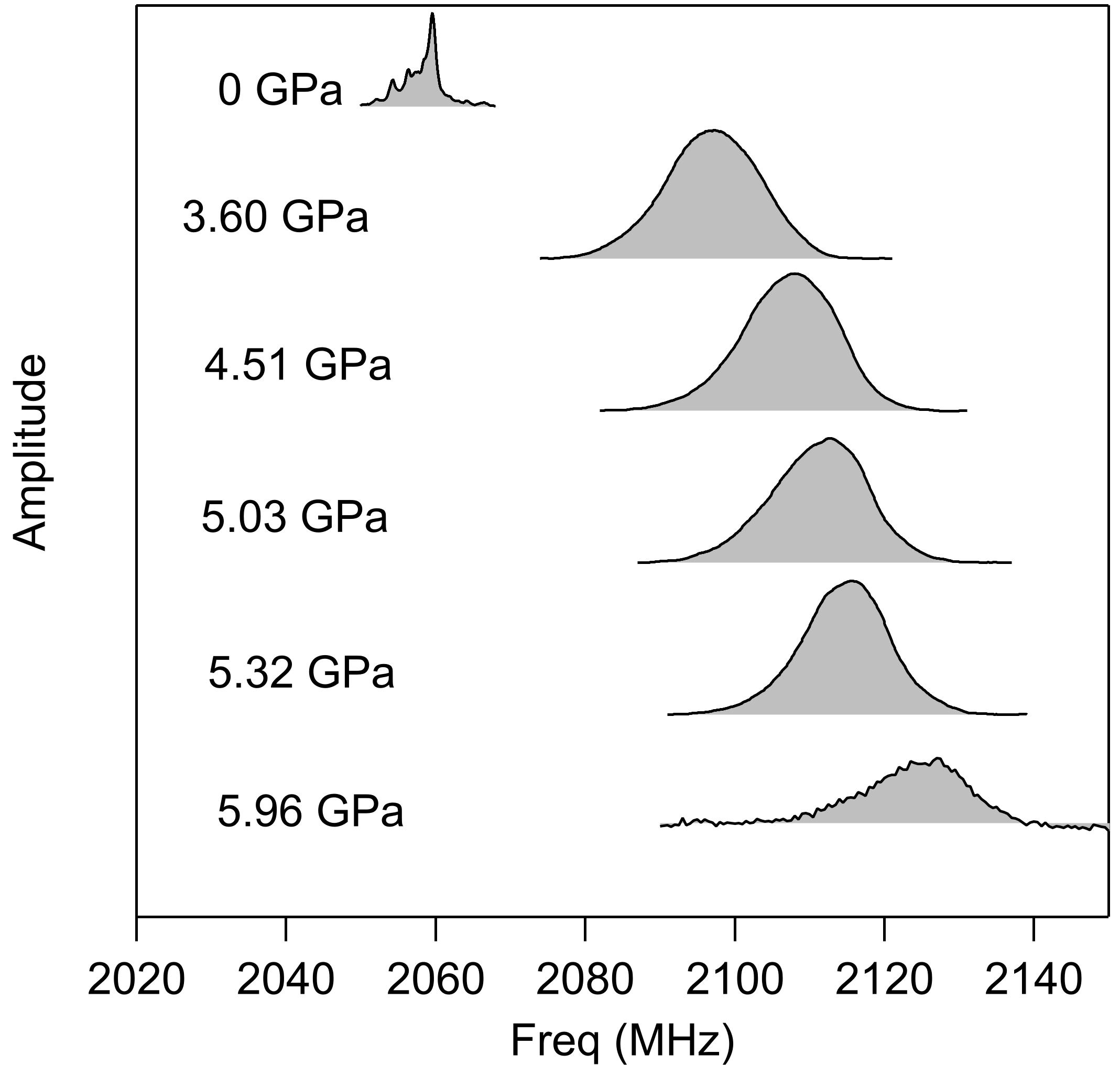}
\caption{Spectra of the $|0\rangle \leftrightarrow |-1\rangle$ transition of the NV ensemble as a function of pressure, measured by echo sequences. The spectrum at 5.96 GPa was measured by continuous wave, since {the antenna leads were thinned out under the pressure from the anvils, reducing the} the Rabi frequency.}
\label{fig:spectra}
\end{center}
\end{figure}

Spectra of the $|0\rangle \leftrightarrow |-1\rangle$ transition under pressure are shown in Fig. \ref{fig:spectra}.  The resonance frequency of this transition is given by $f_0 = D-\gamma_e H_0$, where $D$ is the zero-field splitting, $\gamma_e$ is the electron gyromagnetic ratio, and $H_0 =29$ mT is the magnitude of the applied field.  $D$ increases linearly with pressure \cite{NVpressurePRL,NVinDAC}, which enables us to use the shift of the resonance as a measure of the pressure.  In this case we use the ambient pressure value of $D(0) = 2866$ MHz and $dD/dP = 11.72$ MHz/GPa. The spectra clearly broaden with pressure, and the three peaks are no longer resolvable under pressure. We find that the linewidths increase by a factor of eight over this range, corresponding to a pressure variation of approximately 15\% over the volume of the sample.  The non-hydrostaticity is likely due to freezing of the pressure medium (Daphne oil 7575), which freezes above 3 GPa at room temperature.  Above 5.5 GPa, we found that the Rabi frequency dropped significantly as the leads to the antenna become clipped under pressure.  Utilizing thinner leads in the future may alleviate this problem.

\subsection{Dynamical decoupling}

In the rotating frame, the NV spins will precess around $\mathbf{B}_{ac}(t)$ (Eq. \eqref{eq:1}) and acquire a phase $\Phi(t) = \gamma\int_0^{t}B_{ac}(t')dt'$. A 180$^{\circ}$ pulse will effectively reverse the direction of $\mathbf{B}_{ac}(t)$, and reverse the direction of precession.  For a series of 180$^{\circ}$ pulses spaced at regular intervals, $\tau$, the phase of the NV can be written as \cite{degen2017quantum}:
\begin{equation}
\Phi(t) = \gamma\int_0^{t}B_{eff}(t')dt',
\label{eqn:Phi}
\end{equation}
where the effective field is $B_{eff}(t) = u(t)B_{ac}(t)$, and $u(t)$ alternates between $\pm 1$ between successive 180$^{\circ}$ pulses (which we approximate as infinitely-narrow in time).
The effective field can be expressed as:
\begin{eqnarray}
\nonumber    B_{eff}(t)&=& B_{ac}\sum_{n =1, 3, \cdots} \frac{2}{n\pi} \left[\cos( (\omega_{ac} -{n\pi/\tau})t - \phi_{ac})\right. \\
    &&\left. - \cos( (\omega_{ac} +{n\pi/\tau})t + \phi_{ac}) \right].
    \label{eqn:Beff}
\end{eqnarray}
Most of these frequency components will integrate to zero in Eq. \ref{eqn:Phi}, which reflects that fact that such pulse sequences filter out noise sources with frequency components far from ${n\pi/\tau}$.  On the other hand, if ${n\pi/\tau} \approx \omega_{ac}$, then $B_{eff}(t)$ has a component that is nearly constant over time and the phase $\Phi$ will not integrate to zero.  For $n=1$, where the pulse spacing $\tau\approx \pi/\omega_{ac}$,
\begin{equation}
    \Phi(t) =  \frac{2\gamma B_{ac} t}{\pi}{\mathrm{sinc}}(\pi \delta f t) \cos(\pi\delta f t -\phi_{ac})
\end{equation}
where $\delta f = (2\tau)^{-1} - \omega_{ac}/(2\pi)$  {is the frequency offset}.  If the pulse spacing is tuned exactly such that $\delta f = 0$, then the phase will accumulate linearly with time. After accumulating this phase for a series of $N$ pulses, the NV magnetization can be then be projected to the $z$ axis.  At the end of the sequence, the NV magnetization depends on the relative phase of the first and last $90^{\circ}$ pulses, and that the fluorescence contrast varies as:
\begin{equation}
    \Delta I \sim
    \begin{cases}
        \cos\left(2N\frac{\gamma B_{ac}}{\omega_{ac}}\cos\phi_{ac}\right) & \text{for } 90_x - 90_x\\
         \sin\left(2N\frac{\gamma B_{ac}}{\omega_{ac}}\cos\phi_{ac}\right) & \text{for } 90_x - 90_y\\
    \end{cases}
\end{equation}
For  {typical values of}   {$\omega_{ac}=2\pi \cdot 100$} kHz with amplitude  {$B_{ac}=$}15 nT {and $N=48$ refocusing pulses}, the fluorescence contrast will be reduced by 0.1\%  {for the $90_x-90_y$ sequence, assuming no coherence decay.}

\subsection{Pulse spacing sweep}

\begin{figure}
\begin{center}
\includegraphics[width=\linewidth]{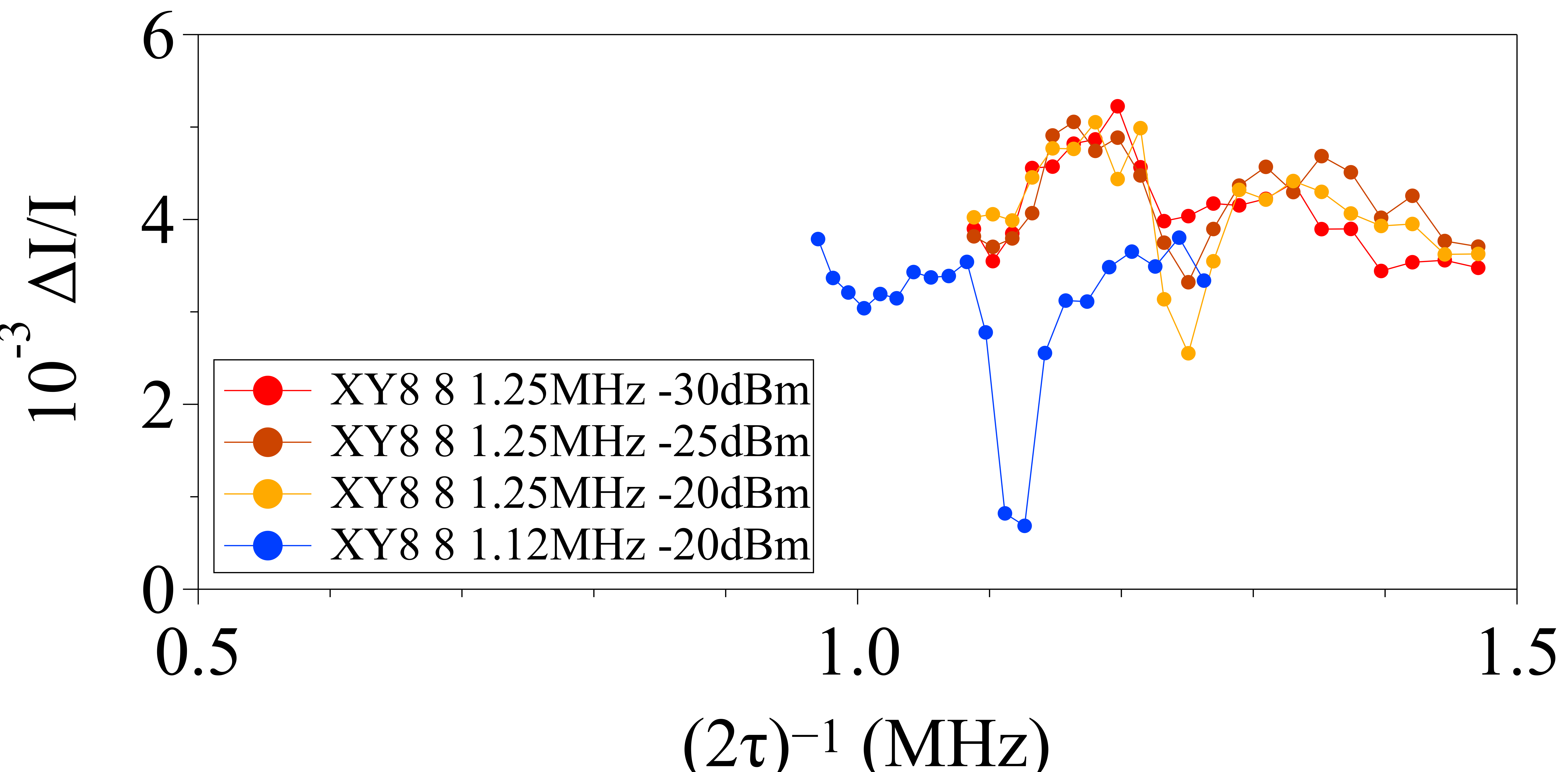}
\includegraphics[width=\linewidth]{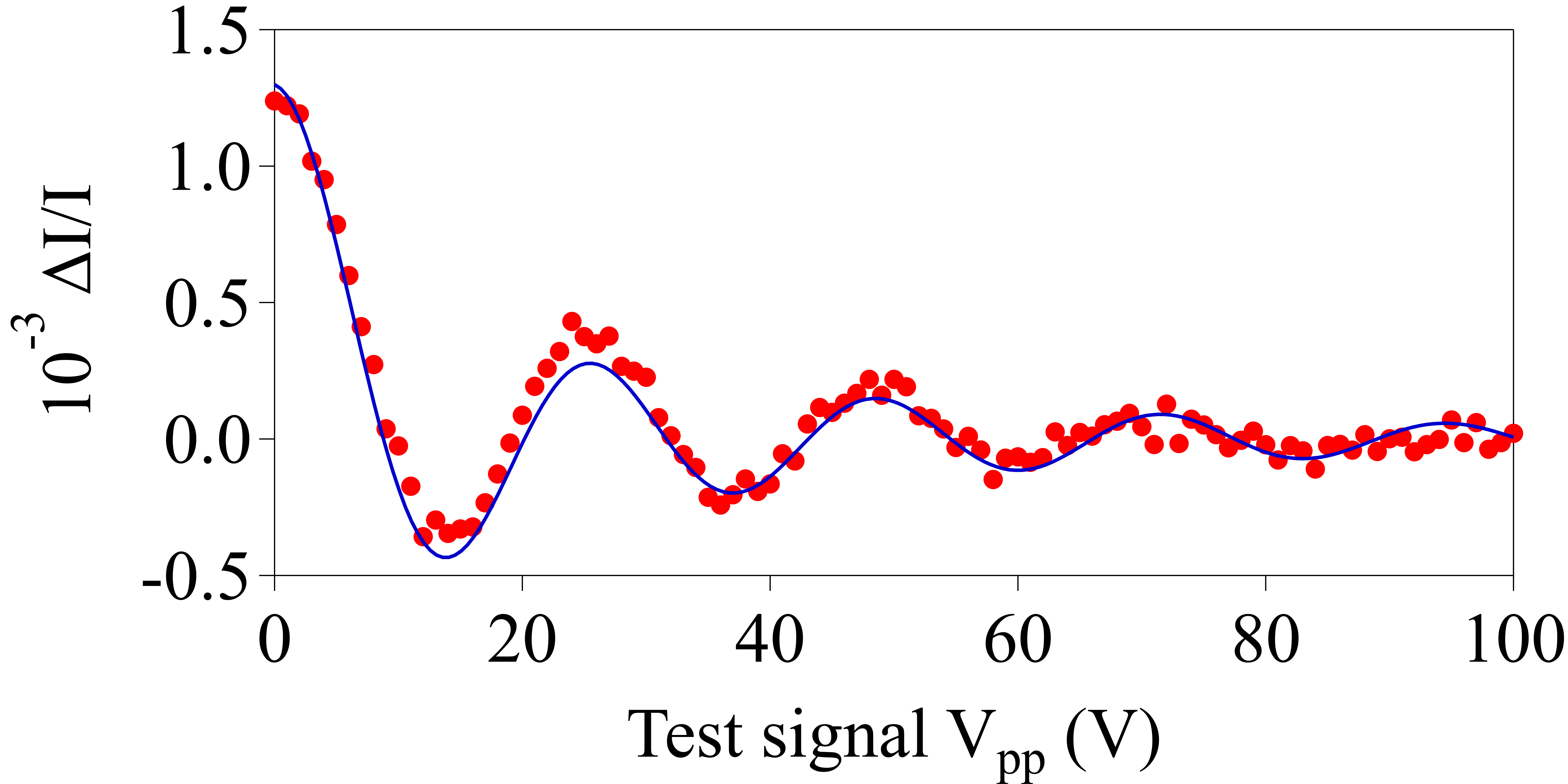}
\caption{(Upper panel) Optical contrast, $\Delta I$ measured in a DAC measured versus detection frequency, $(2\tau)^{-1}$, where $\tau$ is the pulse spacing of a dynamical decoupling sequence.  Input signals of 1.12 and 1.25 MHz were introduced externally. The dip in the contrast arises due to the phase accumulated by the NV centers. (Lower panel) Optical contrast, $\Delta I$ of a test AC signal measured by signal averaging as a function of the voltage applied to the AC coil.  The solid purple line is a fit to a Bessel function as described in the text.  }
\label{fig:ACsweep}
\end{center}
\end{figure}

The most straightforward approach to detecting an AC field is to sweep the pulse spacing, $\tau$, and search for changes in $\Delta I$, as illustrated in Fig. \ref{fig:ACsweep}.  In this case, multiple measurements of the signal contrast are averaged, while a CW external field is applied via a small loop located near the sample space in the diamond anvil cell. The phase, $\phi_{ac}$ is randomly distributed, depending on the time at which the NV detection sequence is initiated.  In this case, for the $90_x-90_x$ sequence, the fluorescence contrast is given by:
\begin{equation}
    \Delta I = 1-J_0(N{\gamma B_{ac}}/{\omega_{ac}}),
\end{equation}
where $J_0(x)$ is a Bessel function of the first kind. For the $90_x-90_y$ sequence $\Delta I = 0$. This behavior is illustrated in the lower panel of Fig. \ref{fig:ACsweep}, in which the AC voltage, $V_{pp}$ is varied to change the magnitude of $B_{ac}$.  In this case, $B_{ac} \propto V_{pp}$, however the coefficient is unknown.

\begin{figure}
\begin{center}
\includegraphics[width=\linewidth]{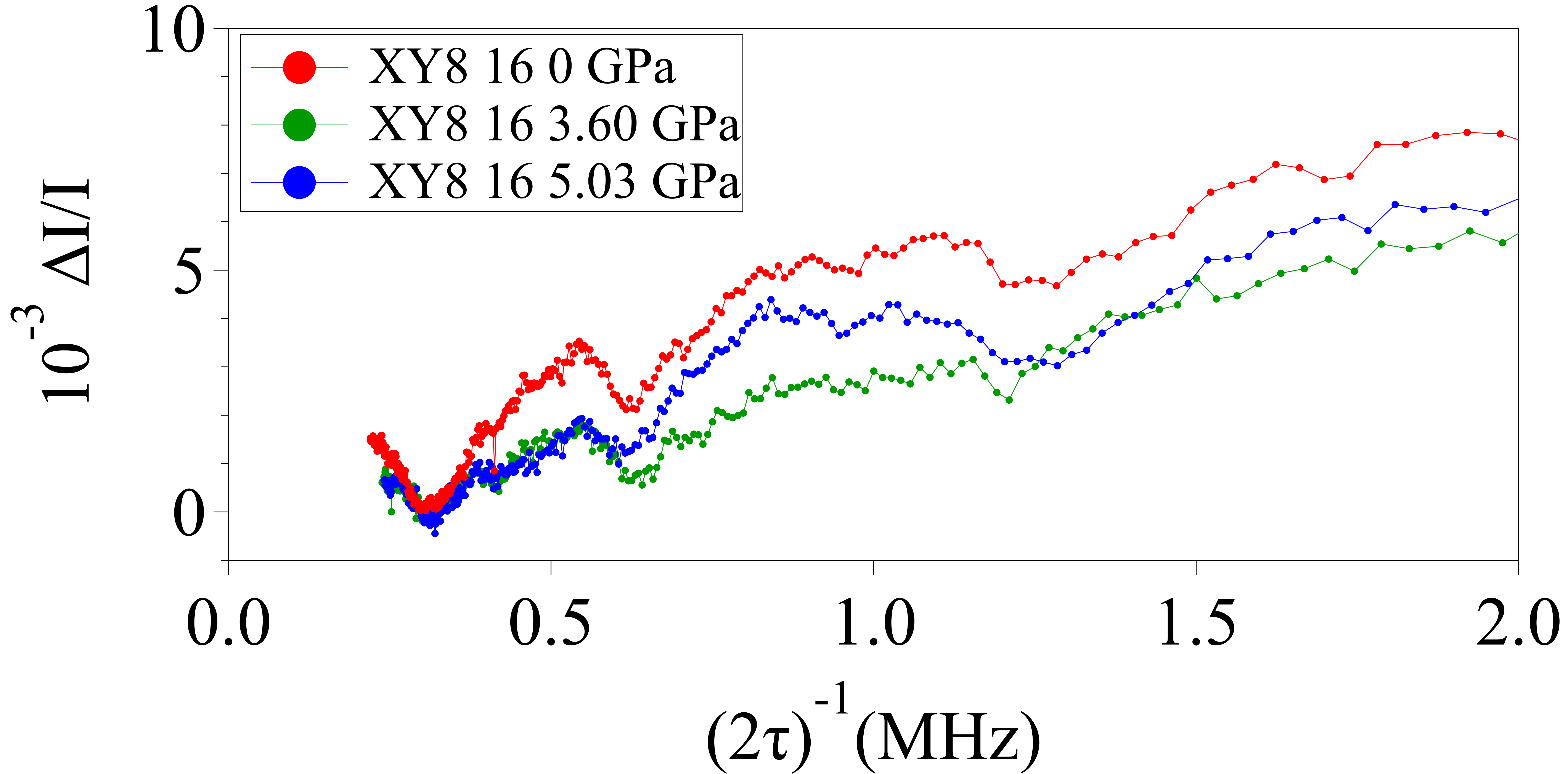}
\caption{Optical contrast versus frequency, $(2\tau)^{-1}$, where $\tau$ is the pulse spacing of an XY8-16 sequence, measured at various pressures.  The dip {s around 0.3 MHz and 1.2 MHz} arises from precessing $^{13}$C spins in the diamond lattice  {and $^1$H spins in the pressure medium, respectively}.}
\label{fig:XY8comparison}
\end{center}
\end{figure}

Figure \ref{fig:XY8comparison} shows AC spectra in the absence of an external driving AC field at ambient pressure, 3.6 GPa and 5.0 GPa respectively.  There is a dip in the contrast at $\sim 0.3$ MHz, which corresponds to the Larmor frequency of $^{13}$C in the applied $H_0$ field.  The 1\% abundant $^{13}$C nuclear spins in the diamond lattice precess at this frequency and create AC fields at the NV site that are not refocused by the dynamical decoupling.  {The dip around 1.2 MHz is likely due to hydrogen in the pressure medium.}

\subsection{Synchronized Readout}

\begin{figure*}
\begin{center}
\includegraphics[width=\linewidth]{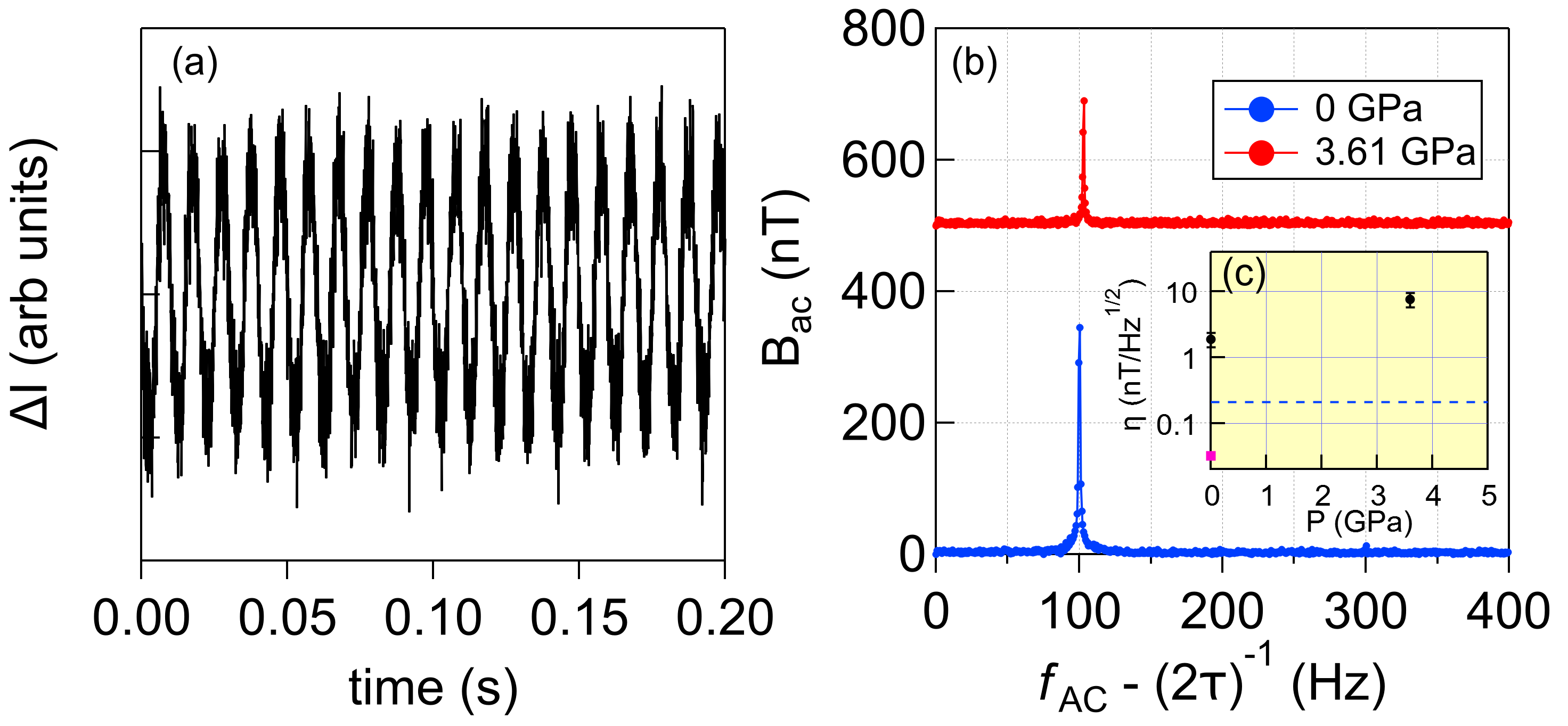}
\caption{(a) Optical contrast, $\Delta I$ measured for NVs in the DAC with a test signal at {frequency $f_{ac} = 1.2001$ MHz, where $\tau = 416.667$ ns (corresponding to a detection frequency at 1.2 MHz)}, for a sequence  {of} equally-spaced measurements at with $T_{meas} = 45$ $\mu$s, as described in the text.   (b) Fourier transform of $\Delta I$ at ambient pressure.  The spectra exhibits a single large peak at 100 Hz, and the vertical axis has been scaled to match the known applied field magnitude. The 3.61 GPa data has been offset vertically by 500 nT for clarity.  {(c) Sensitivity, $\eta$ ($\bullet$), versus pressure.  The dashed blue line corresponds to the necessary sensitivity to achieve unity signal to noise ratio after 1 minute, and the pink square ($\blacksquare$) at 0 GPa corresponds to the sensitivity reported in \cite{Glenn2018}.} }
\label{fig:SRtests}
\end{center}
\end{figure*}

Instead of averaging the signal over randomly-distributed initial phases at the beginning of each repeat of the dynamical decoupling sequence, it is possible to collect each fluorescence measurement and perform a Fourier transform.  After each NV fluorescence measurement, the NV is re-initialized into the ground state without affecting the AC source.  As a result, the phase of $\mathbf{B}_{ac}(t)$ will evolve in a coherent fashion for each measurement.  The slowest varying component of the effective field in Eq. \ref{eqn:Beff} will acquire a phase $\phi_m = 2\pi\delta f m T_{meas} - \phi_{ac}$, where $T_{meas}$ is the measurement repeat time, and $m$ is the $m^{th}$ measurement.  In this case, $\Phi_m\approx \frac{2\gamma B_{ac}T_{meas}}{\pi}\cos\phi_m$.  For a $90_x - 90_y$ sequence in the limit of small $B_{ac}$, the fluorescence contrast will vary as:
\begin{equation}
    \Delta I(T) \sim \left(\frac{2N\gamma B_{ac}}{\omega_{ac}}\right)\cos(2\pi\delta f T - \phi_{ac}),
    \label{eqn:SRdependence}
\end{equation}
where $T = mT_{meas}$.  This behavior is illustrated in Fig. \ref{fig:SRtests}  {both at ambient pressure and at 3.6 GPa}.  In effect, the AC field is demodulated by the dynamical decoupling sequence, and imprinted on $\Delta I$.  A Fourier transform of $\Delta I (T)$ thus provides a spectrum of $B_{ac}(t)$ relative to the detection frequency $(2\tau)^{-1}$.  This detection scheme is often referred to as `quantum heterodyne' \cite{Schmitt2017,Norman2020}.  The frequency resolution is determined by the number of points $m$, $T_{meas}$, the stability of the field $\mathbf{H}_0$, and the stability of the microwave pulse timing. The resolution can be as low as a mHz, which is sufficient to resolve many chemical shifts  even at low applied fields \cite{Schmitt2017,Glenn2018}. This technique requires coherently averaging many repeated measurements, but sufficient signal-to-noise ratio can be obtained within a few hours.

In practice, the minimum detectable frequency {, $\omega_{ac}$,} is bounded by $T_2$ because the NV magnetization will decay over during the free precession time, $\tau$, giving rise to another factor of $e^{-\pi/\omega_{ac}T_2}$. For our NV diamond we have $T_2\approx 200 \mu$s, corresponding to a minimum frequency $\sim 10$ kHz.  The maximum detectable frequency is bounded by the shortest possible pulse spacing, which is approximately equal to the 180$^{\circ}$ pulse time, determined by the Rabi frequency.  In our case the maximum detectable frequency $\sim 10$ MHz. Note that there is an upper limit to the heterodyne frequency, $\delta f$, based upon the minimum time between consecutive measurements.  This limit is set by the time to measure the fluorescence, which is usually a few $\mu$s.

\subsection{Sensitivity Measurements}

\begin{figure}
\begin{center}
\includegraphics[width=\linewidth]{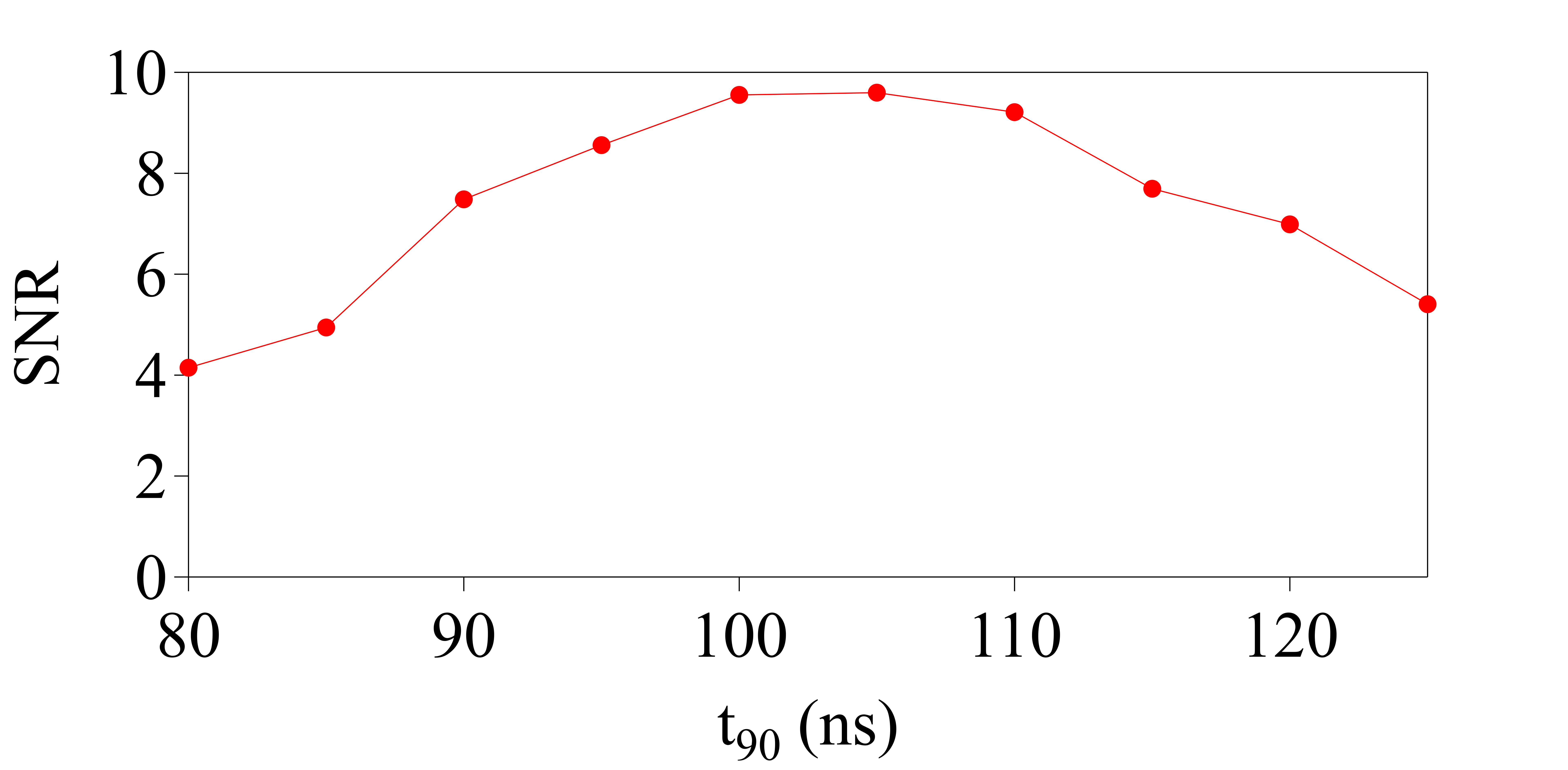}
\caption{Signal intensity (arbitrary units) for a synchronized readout measurement of an AC signal (as in Fig. \ref{fig:SRtests}) versus the width of a 90$^{\circ}$  pulse.  The dynamical decoupling pulses ($180^{\circ}$) are twice as long as the time $t_{90}$ given on the horizontal axis.}
\label{fig:SRvaryt90}
\end{center}
\end{figure}

The data shown in Fig. \ref{fig:SRtests} was acquired for a large applied  {test AC field} for a single acquisition sequence without signal averaging with a total measurement time of approximately 1.5 s, corresponding to a sensitivity,  {$\eta= 1.9$} nT/$\sqrt{\mathrm{Hz}}$  at ambient pressure.  At 3.6 GPa, we find  {the sensitivity decreases to} $\eta =7.6$ nT/$\sqrt{\mathrm{Hz}}$.   {Here we define sensitivity as the minimum detectable signal per unit time. Ideally $\eta$ should be as low as possible to achieve a high signal to noise ratio.} Our values are not as low as previous reports, which reach down to 32 pT/$\sqrt{\mathrm{Hz}}$ \cite{Glenn2018}. One of the primary reasons for the  {difference} in the DAC is the wide distribution of Rabi frequencies, which means that not all of the NVs in the ensemble experience the same $H_1$ field.  The pulse width for the dynamical decoupling 180$^{\circ}$ pulses, given by $\pi/(\gamma H_1)$, will differ for each of these NVs. As a result, the accumulated phase in Eq. \ref{eqn:Phi} is not as high as it would be for a uniform $H_1$. To illustrate this point, Fig. \ref{fig:SRvaryt90} shows how the signal intensity varies as a function of the pulse width time.  Numerical simulations of the  Bloch equations under these conditions indicate that the sensitivity is reduced by $\approx 10$\% if the dynamical decoupling pulses are either 10\% shorter or longer than ideal.  As shown in Fig. \ref{fig:Rabi}, the $H_1$ field for the straight antenna is inhomogeneous over the sample volume, such that the width of the distribution is approximately 40\% of the average.  Although this antenna is able to provide large $H_1$ fields within the sample space, the field inhomogeneity reduces the synchronized readout sensitivity.

Another limiting factor for the sensitivity is the large ratio of the 180$^{\circ}$ pulse width, $t_{180}$, to the pulse spacing, $\tau$.  {The data in Fig. \ref{fig:SRtests} utilized $t_{180}/\tau\approx 0.43$}, which is far from the idealized case of the infinitely-narrow pulses that would give rise to the effective field in Eq. \ref{eqn:Beff}. Numerical simulations indicate that the sensitivity will be suppressed when the ratio $t_{180}/\tau \gtrsim 0.2$.  This provides an effective upper limit on $\omega_{ac}^{max} \approx 0.2\Omega$ for a given value of the Rabi frequency, $\Omega$.

The lower limit for the sensitivity is determined by $T_2$ because the NV magnetization will decay during the free precession time between the dynamical decoupling pulses.  This leads to a exponential prefactor of $e^{-\pi/\omega_{ac} T_2}$ for Eq. \ref{eqn:SRdependence}, and an effective lower limit $\omega_{ac}^{min} \approx \pi/\ln(10) T_2$, where the sensitivity is reduced by a factor of 10.  For our NV sample with $T_2\approx 200$ $\mu$s and $\Omega\approx 3$ MHz, we have $f_{ac}^{min}\approx 1$ kHz and  $f_{ac}^{max}\approx 0.5$ MHz.  For detection of precessing nuclear spins such as $^{1}H$ or $^{19}$F, $H_0$ should lie between 0.02 and 25 mT in order to achieve $\omega_{ac}$ in this range. Curiously, except for small frequencies where $T_2$ limits the sensitivity, the fluorescence contrast maximum in Eq. \ref{eqn:SRdependence} does not depend on the applied field, $H_0$, because both  $B_{ac}$ and $\omega_{ac}$ are proportional to $H_0$ (for thermally polarized nuclei).

 {Under pressure we find that $\eta$ increases by a factor of four, as shown in Fig. \ref{fig:SRtests}(c). There are two possible reasons for this change. Under pressure the NV spectrum is broadened due to the pressure gradients.  If the spectrum is sufficiently broad, the dynamical decoupling pulses will not fully refocus the NV magnetization, similar to the effect of a non-uniform $H_1$ field as discussed above.} {Moreover, the antenna deforms under pressure, and as a result the impedance changes and affects the magnitude of the $H_1$ field.  Consequently both $\Omega$ and $\Delta I$ can be reduced, affecting the sensitivity.}


\section{Discussion and Conclusions}

We have demonstrated AC-ODMR in a DAC with a sensitivity of 1.9 nT$/\sqrt{\mathrm{Hz}}$, which is about 50 times larger than values reported for other NV-based AC sensors operating under ambient conditions. To detect nuclear spins external to the diamond, it is possible to either initialize these with a 90$^{\circ}$ pulse at the nuclear Larmor frequency ($\sim 1$ MHz) in order to let them precess coherently, or to use correlation spectroscopy to detect their noise fluctuations.    {The magnitude of $B_{ac}$ can be estimated assuming that the NV diamond chip is embedded in a spherical volume of uniform nuclear magnetization, $\mathbf{M}(t)$, that is precessing at Larmor frequency $\omega_{ac}$.  In this case, $B_{ac} = \frac{8\pi}{3} \chi_N H_0$, where $\chi_N = {\rho\gamma_n^2\hbar^2 I(I+1)}/{3 k_B T}$ is the Curie susceptibility of nuclei with spin $I$,  and number density $\rho$. For hydrogen in water at room temperature in a field of 10 mT, $B_{ac} \approx 27$ pT.  To achieve a signal to noise ratio of unity with our demonstrated sensitivity $\eta = 1.9$ nT$/\sqrt{\mathrm{Hz}}$ would require signal averaging for a time $T_{avg} = (\eta/B_{ac})^2 \sim 80$ min.  The dashed blue line in Fig. \ref{fig:SRtests}(c) shows the maximum $\eta$ necessary to achieve a minimum signal to noise ratio of unity within one minute.  It is clear that an order of magnitude reduction in $\eta$ will be necessary to avoid hours of signal averaging}.
 {Note that although conventional NMR has sensitivities on the order of 10 pT/$\sqrt{\mathrm{Hz}}$ \cite{degen2017quantum}, this quantity scales as $\eta_{NMR}\sim V_s^{-1/2}$, where $V_s$ is the sample volume \cite{Hoult1976}.  For a 100 $\mu$m coil diameter and sample volume of 1 nL at a field of 10 mT, we estimate $\eta_{NMR} \approx 85$ nT/$\sqrt{\mathrm{Hz}}$ \cite{peck1995design}, approximately two orders of magnitude higher than our measurement.}

 {There are several possible routes to improve the sensitivity of AC-ODMR in the DAC. The  inhomogeneity of the microwave fields from the antenna leads to an increase in $\eta$.  An antenna design that provides a more homogeneous and stable $H_1$ field and is less prone to distortions under pressure is vital.}   In the future other antenna designs may be considered, for example a circular loop within the sample space, to improve the homogeneity and reduce screening effects by the gasket.
Utilizing diamonds with lower NV concentrations and isotopically pure $^{12}$C would significantly enhance $T_2$ and also improve the sensitivity. It is also possible to improve the sensitivity by utilizing dynamic nuclear polarization.   This number can be enhanced significantly by utilizing the presence of free radicals and taking advantage of the Overhauser effect to transfer polarization to the nuclear spins in a liquid.  Bucher et al. used the TEMPOL molecule dissolved into liquid solutions to enhance the sensitivity by more than a factor of 200 \cite{Bucher2020}.  The unpaired electrons on the TEMPOL molecule have an enhanced spin polarization that can be transferred to hyperpolarize the nuclear spins beyond their thermal polarization described by $\chi_N$.  This process can enhance the signal-to-noise ratio by several orders of magnitude, but requires pulsing the system at the Larmor frequency of the TEMPOL electron spins for several ms prior to detection of the nuclear spins with NV magnetometry.

 {The sensitivity may also be improved by utilizing double quantum coherence under pressure \cite{Mamin2014, Marshall2021}}. The AC-ODMR experiments described here are based on single quantum coherence in which the NV is prepared in the state $|0\rangle + |\pm1\rangle$.   {The resonance frequency of the double quantum coherent state $|-1\rangle + |+1\rangle$  is independent of $D$}.  {This {property could be } important because pressure inhomogeneities strongly affect $D$, giving rise to spectral broadening that impacts the sensitivity of the AC detection.}  {Utilizing dynamical decoupling at the double quantum coherent frequency would allow the NV ensemble to precess and acquire phase more homogeneously and may enable a reduction in $\eta$ even in the presence of strong pressure gradients.  It may also be prudent to utilize shaped pulses with the arbitrary waveform generator, which have a broader frequency response and may be able to better refocus the NV magnetization in the presence of inhomogeneous broadening \cite{Spindler2017}.}


 {In addition to the zero-field splitting, $D$, the zero-phonon emission line (637 nm at ambient pressure and temperature) is also  pressure dependent \cite{Doherty20131,NVinDAC}.}  {Both quantities can be used to measure the pressure, eliminating the need for other manometers such as Ruby chips within the sample space  {or the high-frequency edge of the Raman spectra of the diamond anvil} \cite{syassenRubyPress2008,Akahama2006}.  For ODMR, the microwave frequency needs to be adjusted after each pressure change (by 100 GPa it will have increased by 1.2 GHz). This shift may necessitate adjusting different microwave components, such as the amplifier and filters.}  The emission wavelength shifts down with pressure and reaches 532 nm at approximately 60 GPa \cite{NVpressurePRL}. To operate at higher pressures it is thus necessary to switch to a lower wavelength excitation laser.   {It is common to utilize an optical filter before the detector, which needs to be adjusted with increasing pressures.}

AC-ODMR under pressure will enable a broad range of NMR experiments at pressures that have not been possible previously. In particular it should be possible to to test geochemical predictions about fluid chemistry modifying the Earth's crust.  These predictions extend now to 6.0 GPa and 1200 C \cite{sverjensky2014water,sverjensky2015diamond,pan2013dielectric}, and are testable only via vibrational and optical spectroscopic measurements, and not NMR \cite{facq2014situ,facq2016carbon}.  High-pressure NMR spectroscopy via conventional detection is a familiar tool in inorganic solution chemistry \cite{asano1978activation,van1989activation,drljaca1998activation}, but the pressures are less than $\sim 0.4$ GPa, except for a few  {studies} \cite{jonas1980nuclear,lang1977pressure,de1995nmr,ballard1998high}. As discussed above, diamond-anvil cell (DAC) technology can reach higher pressures, but the sample volumes are too small for NMR detection via Faraday coils, which is why optical detection via AC-ODMR is potentially so valuable, particularly if it is coupled to methods of polarization transfer that enhance the signals.  An additional benefit is that experiments are now possible on solutions that would be dangerous in larger amounts.\\

\section{Acknowledgements}
We thank A. Ajoy, S. Gomez-Diaz, P. Klavins, V. Norman and M. Radulaski for stimulating discussions.  The work was supported by the United States Department of Energy, Office of Basic Energy Sciences, Chemical Sciences, Geosciences and Biosciences Division for Grant DE-FG0205ER15693. The data that support the findings of this study are available from the corresponding author upon reasonable request.

\bibliography{ODMRpressure}

\end{document}